\documentclass[12pt]{article}
\usepackage{amsmath}
\usepackage{amssymb}
\usepackage{graphicx}
\usepackage{float}
\usepackage{indentfirst}
\usepackage{mathrsfs}

\usepackage[labelfont=bf,labelsep=period]{caption}

\usepackage[numbers,sort&compress]{natbib}

\newtheorem{theorem}{Theorem}
\newtheorem{proposition}[theorem]{Proposition}

\title{Connection between Measurement Disturbance Relation and  Multipartite Quantum Correlation}

\author
{Jun-Li Li$^{1}$, Kun Du$^{1}$ and Cong-Feng Qiao$^{1,2}$\footnote{corresponding author;
qiaocf@ucas.ac.cn}\\ [0.2cm]
\normalsize{$^1$Department of Physics, University of the Chinese Academy of Sciences,}\\
\normalsize{YuQuan Road 19A, Beijing 100049, China}\\[2pt]
\normalsize{$^2$CAS Center for Excellence in Particle Physics, Beijing 100049, China}
}

\date{}

\begin{document}
\baselineskip24pt \maketitle

\begin{abstract}

It is found that the measurement disturbance relation (MDR) determines the
strength of quantum correlation and hence is one of the essential facets
of the nature of quantum nonlocality. In reverse, the exact form of MDR
may be ascertained through measuring the correlation function. To this
aim, an optical experimental scheme is proposed. Moreover, by virtue of
the correlation function, we find that the quantum entanglement, the
quantum non-locality, and the uncertainty principle can be explicitly
correlated.

\end{abstract}

\section{Introduction}

Quantum nonlocality and Heisenberg's uncertainty principle
\cite{Heisenberg-o} are two essential concepts in quantum mechanics (QM).
The nonclassical information shared among different parts forms the basis of
quantum information and is responsible for many counterintuitive features of
QM, e.g., quantum cryptography \cite{Gisin-2002} and quantum teleportation
\cite{teleporting-1993}. From information theory people have put forward
certain principles to specify the quantum correlation, including nontrivial
communication complexity \cite{communication-complexity-2006}, information
causality \cite{Information-causuality-2009}, entropic uncertainty relations
\cite{entropic-uncertainty-2010}, local orthogonality
\cite{local-orthogonality-2013}, and global exclusivity
\cite{global-exclusivity-C,global-exclusivity-Y}. Note these principles stem
from the notion of information; they mainly concern the bipartite
correlation. It has been shown that understanding multipartite intrinsic
structure is indispensable to the determination of quantum correlation, and
it is rather difficult to derive the Hilbert space structure from
information quantities alone \cite{Require-multi-2011}.

Heisenberg's uncertainty principle has a deep impact on quantum measurement,
and it reflects the mutual influence of  measurement precision and
disturbance on a quantum system only for the measurement disturbance
relation (MDR). The well-known Heisenberg-Robertson uncertainty relation
reads \cite{Robertson}
\begin{eqnarray}
\Delta A\Delta B \geq |\langle C \rangle|
\label{Heisenberg-SD}
\end{eqnarray}
with $C = \frac{1}{2i}[A,B]$ and the standard deviation $\Delta X =
\sqrt{\langle \psi|X^2|\psi \rangle - \langle\psi| X|\psi\rangle^2}$. Note
that in (\ref{Heisenberg-SD}), only the properties of two observables in the
ensemble of a quantum state are involved, and the relation is independent of
any specific measurement. The MDR has been intensively studied both
theoretically \cite{Ozawa-2003-PRA,Hall-2004,
Weston-Pryde,Branciard,MDR-correlation} and experimentally
\cite{Neutron-spin-MDR, Photon-weak-MDR,Ringbauer-2013, Kaneda-2013}.
However, the implication of the MDR for quantum information and quantum
measurement is still unclear
\cite{Busch-2013,Dressel-2013,Quantum-metrology-N}. In practice, there are
different forms of MDR, most of which have undergone experimental checks and
still survive \cite{Ringbauer-2013,Kaneda-2013}. Therefore, determining the
impacts of various MDRs on quantum physics, or ascertaining the right form
of MDR, is currently an urgent task.

In this work, we propose a scheme for transforming any MDR to some
constraint inequalities of multipartite correlation functions. In this way,
the attainable strength of correlations in multipartite state may be
considered to be the physical consequence of the restriction on the quantum
measurement imposed by the MDR. The structure of the paper is as follows. In
Sec. 2, typical versions of the MDR are presented, and their essential
differences are illustrated by embedding the MDR into a coordinate system.
In Sec. 3, we transform the various MDRs to constraint inequalities on
bipartite correlation functions in a tripartite state with the help of the
nonfactorable state. Then it is shown that the constraint inequalities must
be held for all the tripartite and multipartite entangled states. Detailed
examples and an experimental setup for the verification of the various MDRs
based on our scheme are given for a three-qubit system. The concluding
remarks are given in Sec. 4.

\section{The MDR in QM}

\subsection{The quantum measurement and its disturbance}

A quantum measurement process may be generally implemented by coupling a
meter system $|\phi\rangle$ with the original system $|\psi\rangle$. The
measurement result $M$ is obtained from the readout of the meter system. As
the physical observables are represented by Hermitian operators in QM,
following the definition in \cite{Ozawa-2003-PRA}, the measurement precision
of physical quantity $A$ and the corresponding disturbance of quantity $B$
are defined as expectation values of the mean squares:
\begin{eqnarray}
\epsilon(A)^2 \equiv \langle \phi| \langle \psi|
[ \mathcal{A} - A_1 \otimes I_2]^2
|\psi\rangle |\phi\rangle \; , \label{def-precision}\\
\eta(B)^2 \equiv \langle \phi| \langle \psi|
[\mathcal{B}- B_1\otimes I_2]^2|\psi\rangle |\phi\rangle \; . \label{def-disturbance}
\end{eqnarray}
Here $\mathcal{A}=U^{\dag}(I_1\otimes M_2)U$, $\mathcal{B}=U^{\dag}(B_1
\otimes I_2)U$, the subscripts 1 and 2 signify that the operators are acting
on states $|\psi\rangle$ and $|\phi\rangle$, respectively, $U$ is a unitary
interaction between $|\psi\rangle$ and $|\phi\rangle$, and $I$ is the
identity operator. The measurement operator $M$ may be set to $A$ if it has
the same possible outcomes as operator $A$, i.e., $\mathcal{A}=
U^{\dag}(I_1\otimes A_2)U$ \cite{MDR-correlation, Branciard}. Note that in
addition to this operator formalism, which we will work with, there are also
other types of definitions for the measurement precision and disturbance,
e.g., the probability distribution formalism \cite{Busch-2013}.

The MDR indicates that there is a fundamental restriction on the measurement
precision $\epsilon(A)$ and the reaction (disturbance) $\eta(B)$ when two
incompatible physical observables $A$ and $B$ are about to be measured. By
the definitions of (\ref{def-precision}) and (\ref{def-disturbance}),
typical MDR representatives are as follows:
\begin{eqnarray}
& & \epsilon(A)\eta(B)  \geq  |\langle C\rangle| \; , \label{Heisenberg-MDR} \\
& & \epsilon(A)\eta(B) + \epsilon(A)\Delta B +
\eta(B)\Delta A  \geq  |\langle C \rangle| \; , \label{Ozawa-MDR} \\
& & \epsilon(A)\eta(B) + \epsilon(A) \Delta \mathcal{B} +
\eta(B) \Delta\mathcal{A}  \geq  |\langle C\rangle| \; , \label{Hall-MDR}\\
& & \epsilon(A)(\Delta \mathcal{B} + \Delta B) +
\eta(B)(\Delta \mathcal{A} +\Delta A)  \geq  2|\langle C\rangle|
\; , \label{Weston-MDR} \\
& & \Delta B^2 \epsilon(A)^2 + \Delta A^2 \eta(B)^2 +
2\epsilon(A)\eta(B) \sqrt{\Delta A^2\Delta B^2 - \langle C\rangle^2}
\geq \langle C\rangle^2 \; . \label{Branciard-MDR}
\end{eqnarray}
Here $C = [A,B]/2i$, and $\Delta X$ are the standard deviations of operators
$X = A, B, \mathcal{A}, \mathcal{B}$ evaluated in the quantum state
$|\psi\rangle$. Equations (\ref{Heisenberg-MDR})-(\ref{Branciard-MDR})
correspond to Heisenberg-type (He), Ozawa's (Oz) \cite{Ozawa-2003-PRA},
Hall's (Ha)\cite{Hall-2004}, Weston {\it et al}.'s (We) \cite{Weston-Pryde},
and Branciard's (B1) \cite{Branciard} MDRs, respectively. Equation
(\ref{Branciard-MDR}) can be refined, in the specific qubit case, as
\begin{eqnarray}
& & \epsilon(A)^2[1-\epsilon(A)^2/4] +
\eta(B)^2[1-\eta(B)^2/4]
 \nonumber \\ & & + 2\epsilon(A)\eta(B) \sqrt{1-\langle C\rangle^2}
\sqrt{1-\eta(B)^2/4} \sqrt{1-\epsilon(A)^2/4}
 \geq  \langle  C \rangle^2 \; . \label{Branciard-MDR-refine}
\end{eqnarray}
[Equation (9) is abbreviated as B2 bellow.] So far, the Heisenberg-type MDR
has been found to be violated, while others have undergone various sorts of
trials in experiment and still survive \cite{Kaneda-2013,Ringbauer-2013}.
Finding a stricter constraint on $\epsilon(A)$ and $\eta(B)$ is currently a
hot topic in physics \cite{Branciard-13}. Besides focusing on the natures of
different MDRs, it is also important to know what different physical
consequences they would have on quantum information science.

\subsection{The essential difference among various MDRs}

It is obvious that the above representative MDRs differ in tightness. Here
we propose a method for quantitative study of MDRs. We first transform MDRs
into coordinate space and then express them as relation functions of
$\epsilon(A)$ and $\eta(B)$. For the sake of convenience and without loss of
generality, we take three typical MDRs, the Heisenberg type Eq.
(\ref{Heisenberg-MDR}), Ozawa's Eq. (\ref{Ozawa-MDR}), and Branciard's Eq.
(\ref{Branciard-MDR}), as examples.

\begin{figure}\centering
\scalebox{0.25}{\includegraphics{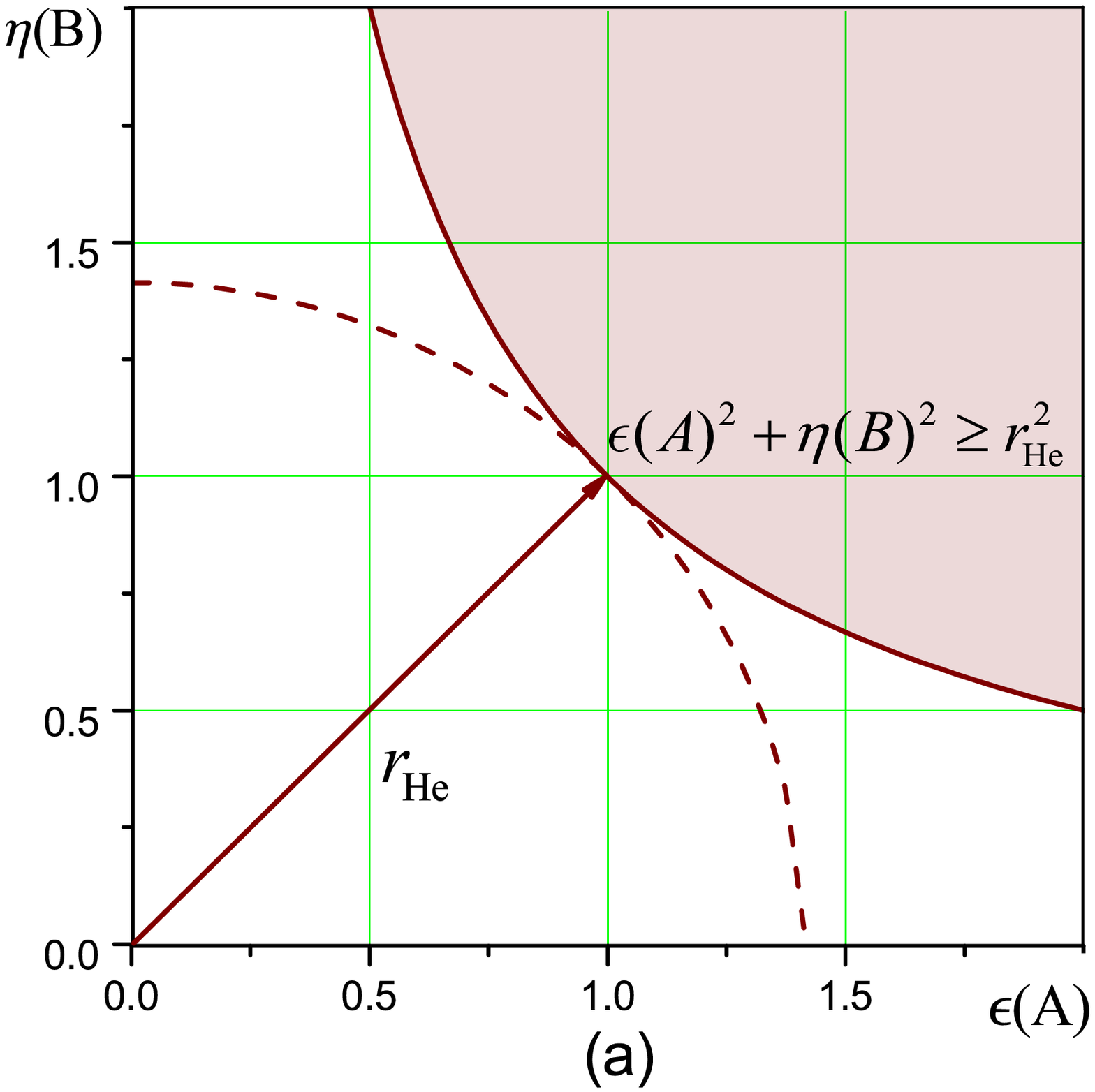}}
\scalebox{0.25}{\includegraphics{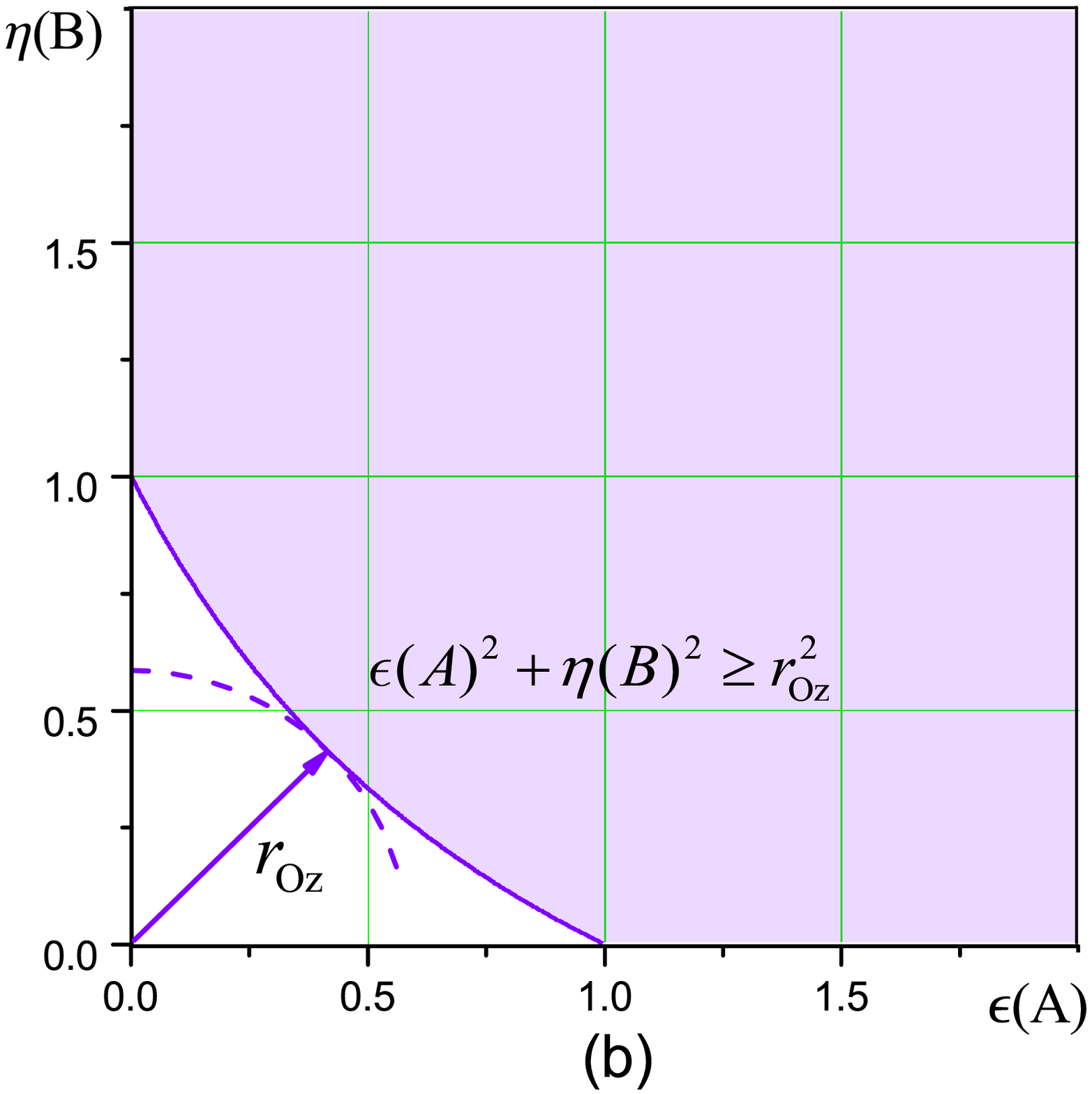}}
\scalebox{0.25}{\includegraphics{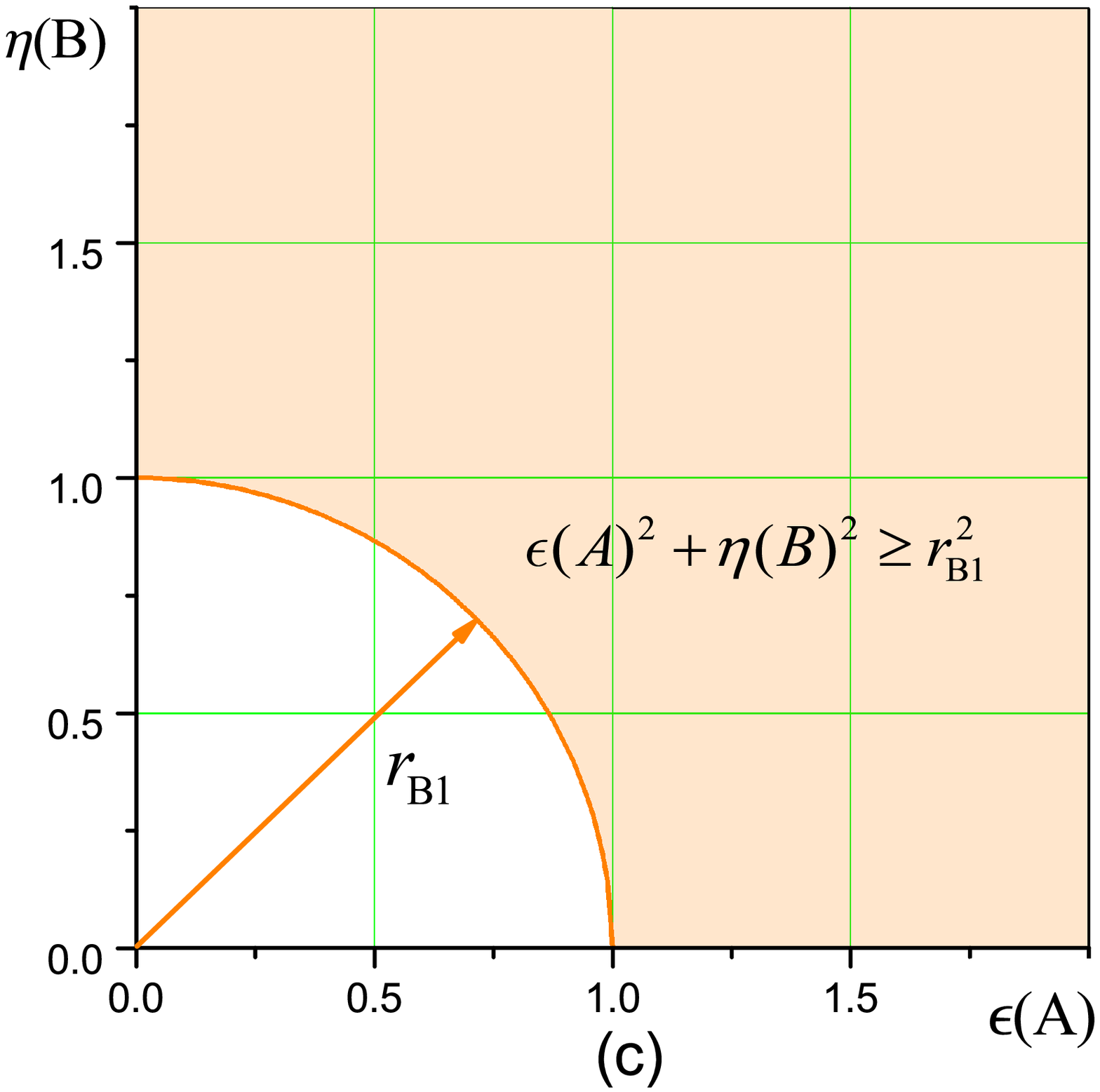}}
\caption{Illustration of different MDRs for the same kinds of
quantum states with identical ensemble properties of $\Delta A$, $\Delta B$, $\langle C\rangle$.
The allowed values of precision $\epsilon(A)$ and disturbance $\eta(B)$ fill the shaded areas,
which correspond to (a) Heisenberg-type, (b) Ozawa's,
and (c) Branciard's MDRs.
The essential differences among those MDRs lie in the forbidden areas for
$\epsilon(A)$ and $\eta(B)$ which are characterized by the minimal
distances to the origin, $r_{\mathrm{He}}$, $r_{\mathrm{Oz}}$,
and $r_{\mathrm{B1}}$, with subscripts stand for the corresponding MDRs.}
\label{Fig-error-circ}
\end{figure}

In the Heisenberg-type MDR (\ref{Heisenberg-MDR}), the measurement-dependent
[$\epsilon(A)$, $\eta(B)$] and measurement-independent ($\langle C\rangle$)
quantities are on different sides of the inequality. The allowed region (AR)
for $\epsilon(A)$ and $\eta(B)$ is above the hyperbolic curve
$\epsilon(A)\eta(B) = |\langle C\rangle|$ in quadrant I [see Fig.
\ref{Fig-error-circ}(a)]. The forbidden region for the values of
$\epsilon(A)$ and $\eta(B)$ is enclosed by the curve and two axes, which may
be characterized by the radius of the circle centered at the origin and
tangent with the hyperbola. This radius represents the minimal distance from
the AR to the origin of the coordinates, $\epsilon(A)^2 + \eta(B)^2 \geq
r_{\mathrm{He}}^2$, where, for the Heisenberg-type MDR, $r_{\mathrm{He}}^2 =
f_{\mathrm{He}}(\langle C \rangle) = 2|\langle C\rangle|$.

For inequality (\ref{Ozawa-MDR}), substituting $ \epsilon(A) =
\epsilon'(A) - \Delta A$ and $\eta(B) = \eta'(B) - \Delta B$, we have
\begin{eqnarray}
\epsilon'(A) \eta'(B) \geq
\Delta A\Delta B + |\langle C\rangle| \; .
\end{eqnarray}
This is a displaced hyperbola of the Heisenberg type [see Fig.
\ref{Fig-error-circ}(b)]. The AR for $\epsilon(A)$ and $\eta(B)$ may be
obtained from (10), and its minimal distance to the origin can also be
expressed as
\begin{eqnarray}
\epsilon(A)^2 + \eta(B)^2 \geq r_{\mathrm{Oz}}^2 = f_{\mathrm{Oz}}(\Delta A,\Delta B,
\langle C\rangle) \; . \label{APP-A-separateform}
\end{eqnarray}

Inequality (8) can be reformulated as
\begin{eqnarray}
\begin{pmatrix} \epsilon(A) &
\eta(B) \end{pmatrix} \begin{pmatrix} \Delta B^2 &
\sqrt{\Delta A^2\Delta B^2 - \langle C\rangle ^2}
\\ \sqrt{\Delta A^2\Delta B^2 - \langle C\rangle^2} & \Delta A^2
\end{pmatrix} \begin{pmatrix} \epsilon(A) \\
\eta(B) \end{pmatrix} \geq \langle C\rangle^2 \; . \label{App-ellipse-fun}
\end{eqnarray}
Different from Heisenberg-type and Ozawa's MDRs, (\ref{App-ellipse-fun}) is
an ellipse of $\epsilon(A)$ and $\eta(B)$ centered at the origin. Similar to
(11), in this case the values of $\epsilon(A)$ and $\eta(B)$ in the AR
satisfy
\begin{eqnarray}
\epsilon(A)^2 + \eta(B)^2 \geq r_{\mathrm{B1}}^2 =
f_{\mathrm{B1}}(\Delta A,\Delta B,\langle C\rangle)
\; ,
\end{eqnarray}
where $r_{\mathrm{B1}}$ is the minor axis of the ellipse with regard to the
parametric condition $\Delta A\Delta B \geq |\langle C\rangle|$. For the
convenience of comparison, the MDRs in Eqs. (\ref{Heisenberg-MDR}),
(\ref{Ozawa-MDR}), (\ref{Branciard-MDR}) are shown in Fig.
\ref{Fig-error-circ} with the same values of $\Delta A$, $\Delta B$, and
$\langle C\rangle$. The values of $\epsilon(A)$ and $\eta(B)$ in the AR fill
up the shaded areas, and the unshaded parts are then the forbidden regions.

To summarize, all of the MDRs (including those yet to be discovered) have
the shortest distance $r_{\mathrm{q}}$ from their AR to the origin as a
function of $\Delta A$, $\Delta B$, and $\langle C\rangle$:
\begin{eqnarray}
\epsilon(A)^2 + \eta(B)^2 \geq r_{\mathrm{q}}^2 =
f_{\mathrm{q}}(\Delta A,\Delta B,\langle C\rangle) \; .
\label{allowed-area}
\end{eqnarray}
Here $f_{\mathrm{q}}$ relies only on the ensemble properties of a
quantum state, i.e., $\Delta A$, $\Delta B$, and $\langle C\rangle$, which
are independent of measurement processes (expressions of $f_{\mathrm{q}}$
for typical MDRs are presented in Appendix {\bf A}). Thus, the shortest
distance $r_{\mathrm{q}}$ from the AR to the origin is independent of the
measurement process and represents the essence of each MDR.

\section{The constraint of MDR on quantum correlation}

\subsection{A nonfactorable bipartite quantum state}

Although variant MDRs may be distinguished by $r_{\mathrm{q}}$, the physical
consequences of different $r_{\mathrm{q}}$ in quantum information theory are
far from obvious. To this end, in this work we present a scheme to examine
MDR. For two Hermite operators $A$ and $B$ with $[A,B]=2iC$, we may
construct a nonfactorable bipartite state $|\psi_{12}\rangle$ satisfying
\begin{eqnarray} A_1\otimes I_2
|\psi_{12}\rangle = I_1 \otimes A'_2 |\psi_{12}\rangle \; ,\; B_1\otimes I_2
|\psi_{12}\rangle = I_1 \otimes B'_2 |\psi_{12}\rangle\; ,
\label{non-factorable12}
\end{eqnarray}
where $A' = UAU^{\dag}$ and $B' = VBV^{\dag}$ are unitary transformations of
$A $ and $B$, respectively, and hence have the same eigenvalues. The
subscripts 1 and 2 indicate the corresponding particles being acted on. We
shall show that for Hermitian operators $A$ and $B$, there always exists
such a nonfactorable state $|\psi_{12}\rangle$.

The Hermitian operators $A$ and $B$ may be expressed in spectrum
decomposition as
\begin{eqnarray}
A  = \sum_{i=1}^{N} \alpha_i|\alpha_i\rangle \langle \alpha_i| \; , \;
B = \sum_{i=1}^{N} \beta_i|\beta_i\rangle \langle \beta_i| \; .
\end{eqnarray}
There is a unitary transformation matrix $W$ between the two orthogonal
bases, $|\beta_i\rangle = \sum_{\mu=1}^{N}|\alpha_{\mu}\rangle w_{\mu i}$,
where $w_{\mu i}$ are the matrix elements of $W$. The following proposition
holds.
\begin{proposition}
If the unitary transformation matrices $U$ and $V$ are congruence
equivalent, that is $U = W V W^{\mathrm{T}}$, then $|\psi_{12}\rangle =
\frac{1}{\sqrt{N}}\sum_{i=1}^{N}|\alpha_i\rangle|\alpha_i'\rangle$ satisfies
\begin{eqnarray}
A_1\otimes I_2 |\psi_{12}\rangle = I_1 \otimes A'_2 |\psi_{12}\rangle \; ,\;
B_1\otimes I_2 |\psi_{12}\rangle = I_1 \otimes B'_2 |\psi_{12}\rangle\;.
\label{supp-proposition-nonfactor}
\end{eqnarray}
Here $A'|\alpha_i'\rangle = \alpha_i|\alpha_i'\rangle$, and the subscripts
of the operators stand for the particles they act on.
\label{proposition-psi}
\end{proposition}
\noindent{\bf Proof:} Given that the bipartite state $|\psi_{12}\rangle =
\frac{1}{\sqrt{N}} \sum_{i=1}^{N} |\alpha_i\rangle|\alpha_i'\rangle$
satisfies the first equality of Eq. (\ref{supp-proposition-nonfactor}),  we
need to prove that the second equality of Eq.
(\ref{supp-proposition-nonfactor}) is also satisfied. The state
$|\psi_{12}\rangle$ may be expressed in the basis of $|\beta_i\rangle$ and
$|\beta_i'\rangle$ as
\begin{eqnarray}
|\psi_{12}\rangle = \sum_{i,j=1}^{N} \gamma_{ij}^{(b)} |\beta_i\rangle |\beta'_j\rangle
 \; , \label{App-B-0}
\end{eqnarray}
where $|\beta_i\rangle$, $|\beta_i'\rangle$ are the eigenvectors of $B$,
$B'$ with the same eigenvalue $\beta_i$, $\gamma_{ij}^{(b)} \in \mathbb{C}$.
We have
\begin{eqnarray}
\left. \begin{array}{l}
|\alpha_i\rangle = \sum_{j}|\alpha'_j\rangle u^{\dag}_{ji} \\
|\beta_i\rangle = \sum_{j} |\alpha_j\rangle w_{ji} \\
|\beta_i'\rangle = \sum_{j} |\beta_j\rangle v_{ji}
\end{array} \right\} \Rightarrow |\beta'_i\rangle =
\sum_{j} \left[ \sum_{k}
\left( \sum_{\nu} |\alpha'_{\nu}\rangle u^{\dag}_{\nu k}
\right) w_{k j} \right]v_{ji} \; ,
\end{eqnarray}
or, more succinctly, $|\beta'_i\rangle = \sum_{j,k ,\nu} u^{\dag}_{\nu k}
w_{k j} v_{ji} |\alpha'_{\nu}\rangle$ with $v_{ji}$, and $u_{\nu k}^{\dag}$
being matrix elements of $V$ and $U^{\dag}$. Therefore, $|\psi_{12}\rangle$
may also be expressed as
\begin{eqnarray}
|\psi_{12}\rangle & = & \sum_{il} \gamma^{(b)}_{il}|\beta_i\rangle |\beta'_{l}\rangle
=  \sum_{i,j,k,l,\mu,\nu}
w_{\mu i} u^{\dag}_{\nu k} w_{k j} v_{jl}
\gamma^{(b)}_{il}|\alpha_{\mu}\rangle |\alpha'_{\nu}\rangle \nonumber \\
& = & \sum_{\mu\nu}\gamma^{(a)}_{\mu\nu} |\alpha_{\mu}\rangle|\alpha_{\nu}'\rangle  \; ,
\end{eqnarray}
where
\begin{eqnarray}
\sum_{i,j,k,l} w_{\mu i} u^{\dag}_{\nu k} w_{k j}
v_{jl}\gamma_{il}^{(b)} = U^{\dag} W V \Gamma^{(b)\mathrm{T}}
W^{\mathrm{T}} = \Gamma^{(a)\mathrm{T}}\; .
\end{eqnarray}
Here $\gamma^{(a)}_{ij}$, $\gamma^{(b)}_{ij}$ are the matrix elements of
$\Gamma^{(a)}$, $\Gamma^{(b)}$ and the superscript $\mathrm{T}$ is the
transpose of a matrix. Because $|\psi_{12}\rangle =
\frac{1}{\sqrt{N}}\sum_{i} |\alpha_i\rangle|\alpha_i'\rangle$, we have
$\gamma_{ij}^{(a)} = \delta_{ij}/\sqrt{N}$, and
\begin{eqnarray}
\Gamma^{(b)\mathrm{T}} & = & V^{\dag}W^{\dag} U \Gamma^{(a)\mathrm{T}} W^{*}
= \frac{1}{\sqrt{N}} V^{\dag}W^{\dag} U W^{*} \nonumber \\
& = &  \frac{1}{\sqrt{N}} V^{\dag}W^{\dag} WVW^{\mathrm{T}}W^{*} =\frac{1}{\sqrt{N}} \; ,
\end{eqnarray}
where the congruence relation $U = W V W^{\mathrm{T}}$ is employed. That is,
\begin{eqnarray}
|\psi_{12}\rangle = \frac{1}{\sqrt{N}} \sum_{i}^{N}|\alpha_i\rangle |\alpha'_i\rangle
= \frac{1}{\sqrt{N}} \sum_i^{N} |\beta_i\rangle |\beta'_i\rangle\; , \label{App-psi12}
\end{eqnarray}
and therefore,
\begin{eqnarray}
A_1\otimes I_2 |\psi_{12}\rangle = I_1 \otimes A'_2
|\psi_{12}\rangle \; ,\; B_1\otimes I_2 |\psi_{12}\rangle = I_1 \otimes B'_2
|\psi_{12}\rangle\;.
\end{eqnarray}
Q.E.D.

Proposition \ref{proposition-psi} indicates that the state
$|\psi_{12}\rangle = \frac{1}{\sqrt{N}} \sum_{i=1}^{N} |\alpha_i\rangle
|\alpha\,'_i\rangle$ satisfies both equalities of Eq.
(\ref{non-factorable12}) while the transformation matrices $U$ and $V$
satisfy $U=WVW^{T}$.

\subsection{The constraint of MDR on quantum correlation}

With the nonfactorable bipartite quantum state in Eq.
(\ref{non-factorable12}), we have the following theorem.
\begin{theorem}
A set of tripartite states may be obtained by an interaction $U_{13}$ of
particle 1 in the state $|\psi_{12}\rangle$ with a third particle (particle
3), $\Psi = \{|\psi_{123}\rangle|\, |\psi_{123}\rangle =
U_{13}|\psi_{12}\rangle|\phi_3\rangle, U_{13}^{\dag}U_{13}=I\}$. The various
MDRs imply the following relationship for states $|\psi_{123}\rangle\in
\Psi$:
\begin{eqnarray}
E(A'_2,A_3) + E(B'_2, B_1) \leq \frac{1}{2}( \langle A_2^{'2} \rangle  +
\langle A_3^2 \rangle + \langle B_2^{'2} \rangle + \langle B_1^2 \rangle
- \gamma_{\mathrm{q}}) \; . \label{Corr-ineq-3}
\end{eqnarray}
Here $E(X_i,Y_j) = \langle X_i \otimes Y_j\rangle$ is the bipartite
correlation function with $X,Y \in \{A,A',B,B'\}$, and $\gamma_{\mathrm{q}}
= \mathrm{Max} \{ \sum_{i} |\ _2\langle p_i|\psi_{12}\rangle|^2
f_{\mathrm{q}}^{(i)}\}$ is independent of $U_{13}$ where $|p_i\rangle$ is an
arbitrary set of orthogonal bases; $f_{\mathrm{q}}^{(i)}$ represents the
function $f_{\mathrm{q}}$ evaluated under quantum states
$|\psi_1^{(i)}\rangle = \ _2\langle p_i|\psi_{12}\rangle/|\ _2\langle
p_i|\psi_{12}\rangle|$. \label{Theorem-MDR-Correlation}
\end{theorem}

\noindent{\bf Proof:} A set of quantum states $|\psi^{(i)}_1\rangle$ of
particle 1 may be prepared by projecting particle 2 in the bipartite
entangled state $|\psi_{12}\rangle$ proposed in Proposition
\ref{proposition-psi} with a set of complete and orthogonal bases
$|p_i\rangle$:
\begin{eqnarray}
|\psi^{(i)}_1\rangle =
\frac{\ _2\langle p_i|\psi_{12}\rangle}{|_2\langle p_i
|\psi_{12}\rangle|} \; . \label{supp-psi1-def}
\end{eqnarray}
Substituting $|\psi_1^{(i)}\rangle$ in the definition of measurement
precision and disturbance [i.e., Eqs. (2) and (3)], we have
\begin{eqnarray}
\epsilon^{(i)}(A)^2 = \langle \phi_3| \langle \psi_1^{(i)}|
[ \mathcal{A} - A_1 \otimes I_3]^2
|\psi_1^{(i)}\rangle |\phi_3\rangle \; , \label{App-(i)A} \\
\eta^{(i)}(B)^2 = \langle \phi_3| \langle \psi_1^{(i)}|
[\mathcal{B}- B_1\otimes I_3]^2|\psi_1^{(i)}\rangle |\phi_3\rangle \; .
\label{App-(i)B}
\end{eqnarray}
Here $|\phi_3\rangle$ describes the meter system. Further, we may write
\begin{eqnarray} |_2\langle p_i
|\psi_{12}\rangle|^2 \epsilon^{(i)}(A)^2 = \langle \phi_3| \langle
\psi_{12}| [U_{13}^{\dag}(I_1\otimes A_3) U_{13} - A_1 \otimes I_3]^2
P^{(i)}_2|\psi_{12}\rangle |\phi_3\rangle \; , \\
|_2\langle p_i
|\psi_{12}\rangle|^2 \eta^{(i)}(B)^2 = \langle \phi_3| \langle \psi_{12}|
[U_{13}^{\dag}(B_1\otimes I_3) U_{13} - B_1 \otimes I_3]^2
P^{(i)}_2|\psi_{12}\rangle |\phi_3\rangle \; ,
\end{eqnarray}
where $P_2^{(i)} = |p_i\rangle_2\langle p_i|$ is a projecting operator
acting on particle 2. Using the complete relation
$\sum_{i}|p_i\rangle\langle p_i| = 1$,
\begin{eqnarray}
\sum_{i} |_2\langle p_i
|\psi_{12}\rangle|^2 \epsilon^{(i)}(A)^2 =\ \langle \phi_3| \langle \psi_{12}|
[U_{13}^{\dag}(I_1\otimes I_2\otimes A_3) U_{13} - A_1 \otimes I_2 \otimes I_3]^2
|\psi_{12}\rangle |\phi_3\rangle \; , \nonumber \\
\sum_i |_2\langle p_i
|\psi_{12}\rangle|^2 \eta^{(i)}(B)^2 =\ \langle \phi_3| \langle \psi_{12}|
[U_{13}^{\dag}(B_1\otimes I_2\otimes I_3) U_{13} - B_1 \otimes I_2\otimes I_3]^2
|\psi_{12}\rangle |\phi_3\rangle\; . \nonumber
\end{eqnarray}
According to Proposition \ref{proposition-psi},
\begin{eqnarray}
\sum_{i} |_2\langle p_i
|\psi_{12}\rangle|^2 \epsilon^{(i)}(A)^2 = \langle \phi_3|\langle \psi_{12}|
[U_{13}^{\dag}(I_1\otimes I_2\otimes A_3) U_{13} - I_1\otimes A'_2\otimes I_3]^2
|\psi_{12}\rangle |\phi_3\rangle \; ,\\
\sum_i |_2\langle p_i
|\psi_{12}\rangle|^2 \eta^{(i)}(B)^2 = \langle \phi_3|\langle \psi_{12}|
[U_{13}^{\dag}(B_1 \otimes I_2\otimes I_3) U_{13} - I_1\otimes B'_2 \otimes I_3]^2
|\psi_{12}\rangle |\phi_3\rangle\; ,
\end{eqnarray}
which give (note the interaction $U_{13}$ commutes with the operators of
particle 2)
\begin{eqnarray}
\sum_{i} |_2\langle p_i
|\psi_{12}\rangle|^2 \epsilon^{(i)}(A)^2 & = & \langle \psi_{123}|
( A_3 - A'_2)^2 |\psi_{123}\rangle \; , \label{App-epsilonA}\\
 \sum_i |_2\langle p_i
|\psi_{12}\rangle|^2 \eta^{(i)}(B)^2 & = & \langle \psi_{123}|
( B_1  -  B'_2)^2 |\psi_{123}\rangle \; . \label{App-etaB}
\end{eqnarray}
Here $|\psi_{123}\rangle = U_{13}|\psi_{12}\rangle|\phi_3\rangle$.

For each quantum state $|\psi_1^{(i)}\rangle$, every MDR has its own AR
region, and its minimal distance to the origin is
\begin{eqnarray}
\epsilon^{(i)}(A)^2 + \eta^{(i)}(B)^2 \geq f^{(i)}_{\mathrm{q}}
(\Delta A,\Delta B,\langle C\rangle)\; , \label{App-radius}
\end{eqnarray}
where the superscript of $f^{(i)}_{\mathrm{q}}(\Delta A,\Delta B,\langle
C\rangle)$ specifies that the argument of the function is evaluated under
the quantum state $|\psi_1^{(i)}\rangle$ and $\mathrm{q}$ stands for He, Oz,
B1, etc. The sum of Eq. (\ref{App-epsilonA}) and Eq. (\ref{App-etaB}) gives
\begin{eqnarray}
\sum_{i} |_2\langle p_i |\psi_{12}\rangle|^2
\left[\epsilon^{(i)}(A)^2 + \eta^{(i)}(B)^2 \right] =
\langle \psi_{123}|
( A_3 - A'_2)^2 + ( B_1  -  B'_2)^2 |\psi_{123}\rangle \; . \label{app-sum-radii}
\end{eqnarray}
Applying (\ref{App-radius}) to (\ref{app-sum-radii}),
\begin{eqnarray}
& & \langle \psi_{123}|
( A_3 - A'_2)^2 + ( B_1  -  B'_2)^2 |\psi_{123}\rangle \nonumber \\
& = & \sum_{i} |_2\langle p_i
|\psi_{12}\rangle|^2 \left[\epsilon^{(i)}(A)^2 + \eta^{(i)}(B)^2 \right]
\geq F_{\mathrm{q}}(|\psi^{(i)}_1\rangle,\Delta A,\Delta B,\langle C\rangle)
\; . \label{App-C-distance}
\end{eqnarray}
Here $F_{\mathrm{q}}(|\psi^{(i)}_1\rangle,\Delta A,\Delta B,\langle
C\rangle) \equiv \sum_{i} |_2\langle p_i |\psi_{12}\rangle|^2
f_{\mathrm{q}}^{(i)}(\Delta A,\Delta B, \langle C\rangle)$. Note that the
establishment of inequality (\ref{App-C-distance}) depends on Eq.
(\ref{App-radius}) but not on the choice of projection bases $|p_i\rangle$.
Therefore the inequality (\ref{App-C-distance}) should also hold as we
optimize the bases $|p_i\rangle$ to get a maximum value of $F_{q}$, that is,
\begin{eqnarray}
\langle \psi_{123}| ( A_3 - A'_2)^2 + ( B_1  -  B'_2)^2
|\psi_{123}\rangle \geq \gamma_{\mathrm{q}} \;,
\label{quadratic-inequality}
\end{eqnarray}
with $\gamma_{\mathrm{q}} = \mathrm{Max} \{\sum_{i} |_2\langle p_i
|\psi_{12}\rangle|^2 f_{\mathrm{q}}^{(i)}\}$. Expanding the quadratic terms,
Eq. (\ref{quadratic-inequality}) turns into the constraint on correlation
$E(A_3,A_2') + E(B_1,B_2')$,
\begin{eqnarray}
E(A_3,A_2') + E(B_1,B_2') \leq
\frac{1}{2}(\langle A_3^2 \rangle + \langle A_2^{'2} \rangle +
\langle B_1^2 \rangle + \langle B_2^{'2} \rangle - \gamma_{\mathrm{q}}) \; ,
\label{Theorem-1}
\end{eqnarray}
with $E(X_i,Y_j) = \langle \psi_{123}|X_iY_j|\psi_{123}\rangle$ and $X,Y \in
\{A,A',B,B'\}$. Q.E.D.

The theorem may be summarized as follows: (1) When a pair of incompatible
observables $A$, $B$ is given, the bipartite entangled state
$|\psi_{12}\rangle$ exists, and one particle of this state may interact with
a third particle $|\phi_3\rangle$ by $U_{13}$. (2) Different forms of MDR
may give different constraints on the bipartite correlations that can be
shared with a third particle. (3) Each constraint inequality, characterized
by $\gamma_{\mathrm{q}}$, is independent of the interaction $U_{13}$, and is
satisfied by all the quantum states in the set $\Psi$. The constraint in the
form of Eq. (\ref{Corr-ineq-3}) is satisfied by all tripartite pure systems;
for details see discussions on the universality of Theorem
\ref{Theorem-MDR-Correlation} presented in Appendix {\bf B}.

\begin{figure}\centering
\scalebox{0.5}{\includegraphics{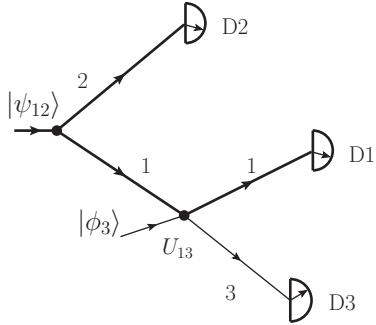}}
\caption{ Schematic illustration of the measurement process of Theorem \ref{Theorem-MDR-Correlation}.
A bipartite state $|\psi_{12}\rangle$ of particles 1 and 2 is prepared, and then
the measurement process is carried out via an interaction with a third particle (particle 3)
 via $U_{13}$. After the interaction, the bipartite correlations are measured for the
resulting tripartite state $|\psi_{123}\rangle = U_{13}|\psi_{12}\rangle|\phi_3\rangle$
at detectors D1, D2, and D3. } \label{App-fig-three-partite}
\end{figure}

The measurement process of Theorem \ref{Theorem-MDR-Correlation} is
illustrated in Fig. \ref{App-fig-three-partite}. According to Theorem
\ref{Theorem-MDR-Correlation}, when one of the entangled particles interacts
with a third particle, the maximal quantum correlation that may be shared
with the third party is not determined by the interaction but by the MDR.
From Eq. (\ref{Corr-ineq-3}) it is clear that the larger the forbidden area
of measurement precision and disturbance is, the less correlation the MDR
predicts. The generalization of Theorem \ref{Theorem-MDR-Correlation} to
incorporate multipartite states may be realized by incorporating successive
measurements with more meter systems. Other generalizations are also
possible as the measurement processes might be implemented in different
scenarios.

\subsection{Physical consequences of MDR in a three-qubit system}

Here we give a detailed example for three-qubit system to show that
different MDRs indeed give different constraints on bipartite quantum
correlations.

\begin{figure}\centering
\scalebox{0.3}{\includegraphics{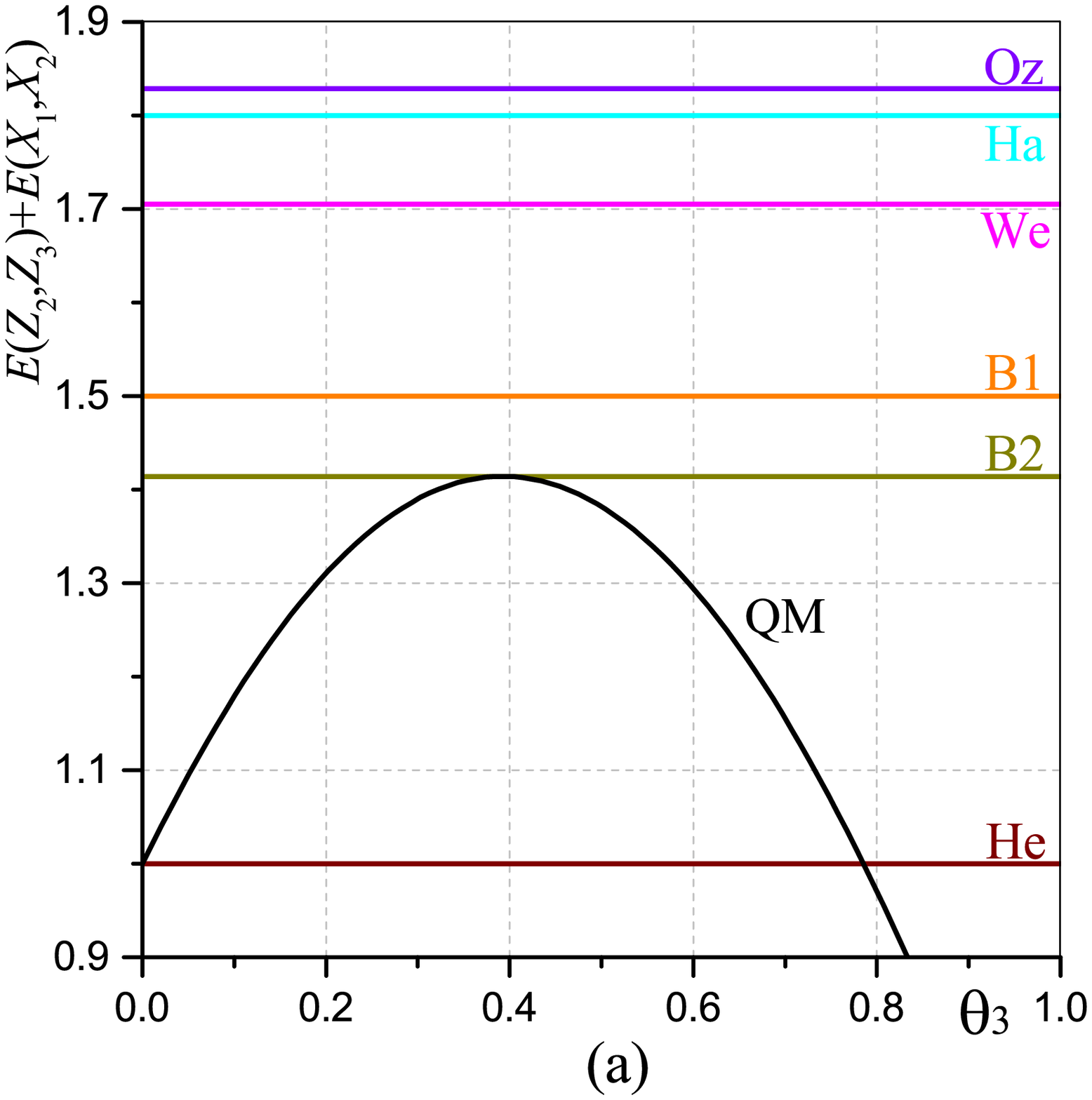}}  \hspace{0.3cm}
\scalebox{0.3}{\includegraphics{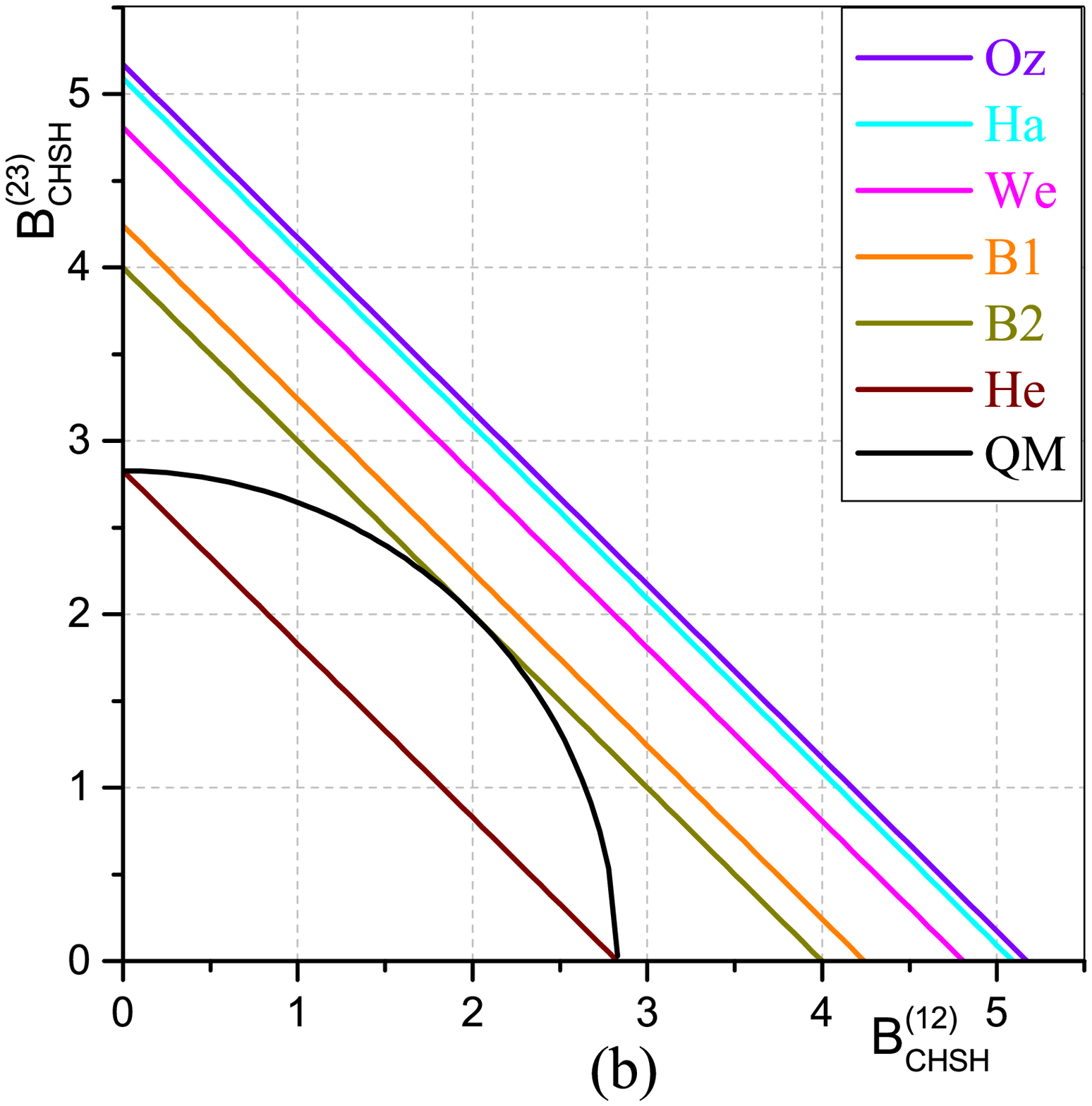}}
\caption{The supremum  for correlation
functions predicted by different MDRs. (a) The supremum
imposed by different MDRs on the sum of bipartite correlation functions
$E(X_1,X_2)$ of particles 1 and 2 and $E(Z_2,Z_3)$ of particle 2 and 3.
(b) Constraints of the sum of CHSH Bell operators for
particles 1 and 2 and 2 and 3 imposed by the MDRs.
Here the abbreviations He, Oz, Ha, We, B1, and
B2 indicate the MDRs of Eqs.
(\ref{Heisenberg-MDR})-(\ref{Branciard-MDR-refine}), respectively. The QM
prediction (black line) contradicts the constraint of the Heisenberg-type MDR.}
\label{Fig-MDRs-qubit-result}
\end{figure}

In a qubit system, two incompatible operators may be set to $A = Z =
\sigma_z$ and $B=X=\sigma_x$. The nonfactorable state can be generally
constructed as
\begin{eqnarray}
|\psi_{12}\rangle = \frac{1}{\sqrt{2}}(|++\rangle + |--\rangle) \; ,
\end{eqnarray}
where $\sigma_z|\pm\rangle=\pm|\pm\rangle$ and we have chosen $A'=A$,
$B'=B$. It is easy to verify
\begin{eqnarray}
A_1|\psi_{12}\rangle = A_2|\psi_{12}\rangle \; , \; B_1|\psi_{12}\rangle
= B_2|\psi_{12}\rangle \; .
\end{eqnarray}
Then let particle 1 interact with particle 3, $|\phi_3\rangle =
\cos\theta_3|+\rangle + \sin\theta_3|-\rangle$, in arbitrary form. Suppose
the interaction is a controlled NOT (CNOT) gate between particles 1 and 3,
\begin{eqnarray}
|\psi_{123}\rangle & = & U_{\mathrm{CNOT}}|\psi_{12}\rangle
(\cos\theta_3|+\rangle + \sin\theta_3|-\rangle) \nonumber \\
& = & \frac{1}{\sqrt{2}}\left[|++\rangle (\cos\theta_3|+\rangle
+ \sin\theta_3|-\rangle)  \right.\nonumber \\
& & \hspace{0.8cm}\left. +|--\rangle (\cos\theta_3|-\rangle
+ \sin\theta_3|+\rangle)\right] \; .\nonumber
\end{eqnarray}
Here particle 1 is the control qubit, and particle 3 is the target qubit.
According to Theorem \ref{Theorem-MDR-Correlation}, we have
\begin{eqnarray}
& & E(A'_2,A_3) + E(B_1,B_2') = E(Z_2,Z_3) + E(X_1,X_2) \nonumber \\
& \leq & \frac{1}{2} (\langle Z_3^2\rangle + \langle Z_2^2\rangle +
\langle X_1^2\rangle + \langle X_2^2\rangle - \gamma_{\mathrm{q}})= 2 -
\frac{\gamma_{\mathrm{q}}}{2} \; , \label{supp-sum-corr-qu}
\end{eqnarray}
under the condition $Z^2=X^2=1$ and with the real parameter
\begin{eqnarray}
\gamma_{\mathrm{q}} = \mathrm{Max}\left[\sum_{i}|\ _2\langle p_i|\psi_{12}\rangle|^2
 f_{\mathrm{q}}^{(i)}\right] \; .
\end{eqnarray}
We may choose any set of complete and orthogonal bases $|p_i\rangle$ to test
the MDR. Generally, we need to optimize the choice of the bases in order to
obtain the maximum value of $\gamma_{\mathrm{q}}$. For a qubit system, by
choosing $|p_{1}\rangle = (|+\rangle +i|-\rangle)/\sqrt{2}$ and
$|p_{2}\rangle = (|+\rangle -i|-\rangle)/\sqrt{2}$, which are the
eigenvectors of $\sigma_y$ with eigenvalues of $+1$ and $-1$, respectively,
we have
\begin{eqnarray}
|\psi_1^{(1)} \rangle & = & \frac{\ _2\langle p_1|\psi_{12}\rangle}{|\ _2\langle p_1|\psi_{12}\rangle|}
= \frac{1}{\sqrt{2}}(|+\rangle - i|-\rangle) \; , \label{state-1} \\
|\psi_1^{(2)} \rangle & = & \frac{\ _2\langle p_2|\psi_{12}\rangle}{|\ _2\langle p_2|\psi_{12}\rangle|}
= \frac{1}{\sqrt{2}}(|+\rangle + i|-\rangle) \label{state-2} \; .
\end{eqnarray}
Because for quantum states $|\psi_{1}^{(i)}\rangle$, [Eqs. (\ref{state-1})
and (\ref{state-2})] we have $\langle \sigma_{z,x} \rangle=0$, $\Delta
\sigma_{z} = \Delta \sigma_{x} = \sqrt{\langle \sigma_y\rangle}$ (see
Appendix {\bf A}), the functions $f_{\mathrm{q}}^{(i)}$ in Eq.
(\ref{App-radius}) reach the maximum. That is
\begin{eqnarray}
f_{\mathrm{q}}^{(1)} = \kappa_{\mathrm{q}}
|\langle \psi_{1}^{(1)}| C |\psi_{1}^{(1)}\rangle| \; , \;
f_{\mathrm{q}}^{(2)} = \kappa_{\mathrm{q}}
|\langle \psi_{1}^{(2)}| C |\psi_{1}^{(2)}\rangle|\; ,
\end{eqnarray}
where $\kappa_{\mathrm{He}} = 2$, $\kappa_{\mathrm{B2}} = (4-2\sqrt{2})$,
$\kappa_{\mathrm{B1}} = 1$, $\kappa_{\mathrm{We}} = 0.59$,
$\kappa_{\mathrm{Ha}} = 2/5$, $\kappa_{\mathrm{Oz}} = (2-\sqrt{2})^2$, and
$C = [A,B]/2i = \sigma_y$. Therefore,
\begin{eqnarray}
\gamma_{q} & = & \left[\sum_{i=1}^{2}
\left|\ _2\langle p_i| \psi_{12}\rangle \right|^2
\kappa_{\mathrm{q}}
\left|\langle \psi^{(i)}_1| C |\psi^{(i)}_1\rangle \right| \right] =  \kappa_{\mathrm{q}} \; .
\end{eqnarray}
Substituting $\gamma_{\mathrm{q}}$ into Eq. (\ref{supp-sum-corr-qu}), we have
\begin{eqnarray}
\text{He}  & : &  E(Z_2,Z_3) + E(X_1,X_2)
\leq 1 \; , \nonumber \\
\text{B2}      & : &  E(Z_2,Z_3) + E(X_1,X_2)
\leq \sqrt{2} \; , \nonumber \\
\text{B1}      & : &  E(Z_2,Z_3) + E(X_1,X_2)
\leq \frac{3}{2}\; , \nonumber \\
\text{We} & : &  E(Z_2,Z_3) + E(X_1,X_2)
\leq 1.71\; , \label{corr-E} \\
\text{Ha}                 & : &  E(Z_2,Z_3) + E(X_1,X_2)
\leq \frac{9}{5}\; , \nonumber \\
\text{Oz}             & : &  E(Z_2,Z_3) + E(X_1,X_2)
\leq 2\sqrt{2}-1\; , \nonumber
\end{eqnarray}
while the QM prediction is
\begin{eqnarray}
E(Z_2, Z_3) + E(X_1, X_2) & = & \langle \psi_{123}|I_1\otimes Z_2\otimes Z_3 |\psi_{123}\rangle
 + \langle \psi_{123}|X_1\otimes X_2\otimes I_3 |\psi_{123}\rangle \nonumber \\
& = & \cos(2\theta_3) + \sin(2\theta_3) \; . \label{App-corr-sum}
\end{eqnarray}
The constraints from MDR and QM on correlation functions are illustrated in
Fig. \ref{Fig-MDRs-qubit-result}(a).

The constraints on correlation functions tend to unveil a more intrinsic
nature of the nonlocal system when we transform them into a
Clauser-Horne-Shimony-Holt (CHSH) Bell inequality \cite{CHSH}. Equation
(\ref{supp-sum-corr-qu}) gives the measurement precision of $Z$ and the
disturbance on $X$ for the qubit system
\begin{eqnarray}
E(Z_2,Z_3) + E(X_1,X_2) \leq  2 - \frac{\gamma_{\mathrm{q}}}{2} \; . \label{supp-CHSH-1}
\end{eqnarray}
Similarly, for the measurement precision of $X$ and the disturbance on $Z$,
we have
\begin{eqnarray}
E(X_2,X_3) + E(Z_1,Z_2) \leq  2 - \frac{\gamma_{\mathrm{q}}}{2} \; . \label{supp-CHSH-2}
\end{eqnarray}
Combining Eq. (\ref{supp-CHSH-1}) with Eq. (\ref{supp-CHSH-2}), we have
\begin{eqnarray}
E(Z_2,Z_3) + E(X_1,X_2) + E(X_2,X_3) + E(Z_1,Z_2) \leq
4 - \gamma_{\mathrm{q}}\; . \label{supp-CHSH-sum}
\end{eqnarray}
Based on the method introduced in Ref. \cite{MDR-correlation}, here we
introduce two additional directions, $\vec{c}=\frac{1}{\sqrt{2}}(1,0,1)$,
$\vec{d}=\frac{1}{\sqrt{2}}(1,0,-1)$, in the real space of the $z$-$x$
plane,
\begin{eqnarray}
\vec{z} = \vec{a} = (0,0,1) = \frac{1}{\sqrt{2}}(\vec{c}-\vec{d}) \; , \;
\vec{x} = \vec{b} = (1,0,0) = \frac{1}{\sqrt{2}}(\vec{c}+\vec{d}) \; .
\end{eqnarray}
Equation (\ref{supp-CHSH-sum}) can then be reexpressed as
\begin{eqnarray}
E(\hat{a}_2,\hat{c}_3) -  E(\hat{a}_2,\hat{d}_3) +
E(\hat{b}_1,\hat{c}_2) +  E(\hat{b}_1,\hat{d}_2)  & & \nonumber \\
+ E(\hat{b}_2,\hat{c}_3) +  E(\hat{b}_2,\hat{d}_3) +
E(\hat{a}_1,\hat{c}_2) -  E(\hat{a}_1,\hat{d}_2) & \leq & \sqrt{2}(
4 - \gamma_{\mathrm{q}}) \; , \label{supp-CHSH-before}
\end{eqnarray}
where $\hat{a}=\vec{\sigma}\cdot \vec{z} = Z$, $\hat{b}=\vec{\sigma}\cdot
\vec{x} = X$, $\hat{c} = \vec{\sigma}\cdot \vec{c}$, $\hat{d} =
\vec{\sigma}\cdot \vec{d}$. Through some rearrangement, Eq.
(\ref{supp-CHSH-before}) now turns into a more transparent form
\begin{eqnarray}
E(\hat{a}_2,\hat{c}_3) -  E(\hat{a}_2,\hat{d}_3) +
E(\hat{b}_2,\hat{c}_3) +  E(\hat{b}_2,\hat{d}_3)  & & \nonumber \\
+ E(\hat{b}_1,\hat{c}_2) +  E(\hat{b}_1,\hat{d}_2) +
E(\hat{a}_1,\hat{c}_2) -  E(\hat{a}_1,\hat{d}_2) & \leq & \sqrt{2}(
4 - \gamma_{\mathrm{q}}) \; .
\end{eqnarray}
This is just the constraint on two CHSH Bell operators for particles 1 and 2
and particles 2 and 3,
\begin{eqnarray}
B_{\mathrm{CHSH}}^{(23)} + B_{\mathrm{CHSH}}^{(12)}
\leq 2\sqrt{2}(2-\frac{\gamma_{\mathrm{q}}}{2}) \; .
\end{eqnarray}
Therefore, for MDRs in the qubit system we have
\begin{eqnarray}
\text{He}   & : &
B_{\mathrm{CHSH}}^{(23)} + B_{\mathrm{CHSH}}^{(12)} \leq 2\sqrt{2} \; , \nonumber \\
\text{B2}        & : &
B_{\mathrm{CHSH}}^{(23)} + B_{\mathrm{CHSH}}^{(12)} \leq 4 \; , \nonumber \\
\text{B1} & : &
B_{\mathrm{CHSH}}^{(23)} + B_{\mathrm{CHSH}}^{(12)} \leq 3\sqrt{2}\; , \nonumber \\
\text{We}       & : &
B_{\mathrm{CHSH}}^{(23)} + B_{\mathrm{CHSH}}^{(12)} \leq 3.42\sqrt{2}\; , \label{Bell-MDR} \\
\text{Ha}                  & : &
B_{\mathrm{CHSH}}^{(23)} + B_{\mathrm{CHSH}}^{(12)} \leq \frac{18\sqrt{2}}{5}\; , \nonumber \\
\text{Oz}                & : &
B_{\mathrm{CHSH}}^{(23)} + B_{\mathrm{CHSH}}^{(12)} \leq 8-2\sqrt{2} \; , \nonumber
\end{eqnarray}
while the QM prediction is \cite{Bell-monogamy2}
\begin{eqnarray}
\langle B_{\mathrm{CHSH}}^{(23)}\rangle^2 +
\langle B_{\mathrm{CHSH}}^{(12)}\rangle^2 \leq 8 \; . \label{CHSH-square}
\end{eqnarray}
The results from the MDR and QM prediction are shown in Fig.
\ref{Fig-MDRs-qubit-result}(b). Here we see that the second Branciard MDR
[Eq. (9)] gives the same supremum as Eq. (\ref{CHSH-square}) on the sum of
two Bell operators.

From Fig. \ref{Fig-MDRs-qubit-result}(a) we notice that, while the supremum
from the Heisenberg-type MDR is 1 in the given configuration, the QM
prediction is $\sqrt{2}$, the largest value for QM (see Appendix {\bf C}).
We conclude that, for every MDR which can be expressed in the operator
formalism, there will be a concrete constraint on the quantum correlations
in the multipartite state. The MDR manifests itself as a principle
determining the strength of the correlations, which may be shared with other
particles through interaction. In the form of the CHSH inequality in Fig.
\ref{Fig-MDRs-qubit-result}(b), the MDR also provides a physical origin of
the monogamy of the entanglement in multipartite entangled states. On the
other hand, the exact form of the monogamy relation for entanglement may
also be used in the reverse to obtain the exact form of MDR.

\subsection{Experimental verification of the MDR}

\begin{figure}\centering
\scalebox{0.5}{\includegraphics{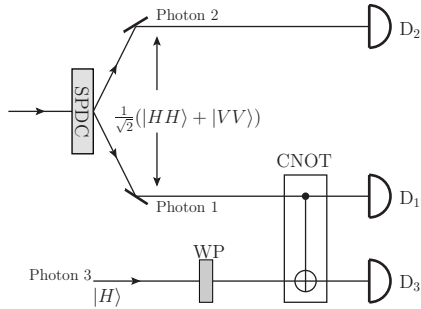}}
\caption{Experimental setup for the verification of MDRs.
A pair of polarization entangled photons (photons 1 and 2) is
generated by means of spontaneous parametric down-conversion (SPDC). A third
photon passing through wave plates (WP), the meter system, interacts with photon 1 via a CNOT operation.}
\label{Fig-MDRs-experiment}
\end{figure}

Except for the fundamental physical implications for multipartite
correlations, the MDR's unique constraint on the bipartite correlation
function in the tripartite state makes the experimental test on MDR
applicable in various physical systems, e.g., atoms, ions, and even higher
energy particles, through measurement of correlation functions
\cite{MDR-correlation}. One schematic optical experimental setup for a qubit
system is shown in Fig. \ref{Fig-MDRs-experiment}. A pair of
polarization-entangled photons, $|\psi_{12}\rangle =
\frac{1}{\sqrt{2}}(|HH\rangle + |VV\rangle)$, is generated by spontaneous
parametric down conversion (SPDC). The meter system of the photon state may
be tuned into the state $|\phi_3\rangle = \cos\theta_3|H\rangle +
\sin\theta_3|V\rangle$ by wave plates (WP). Then it interacts with photon 1
via a CNOT operation resulting in the tripartite state $|\psi_{123}\rangle =
\frac{1}{\sqrt{2}} [|HH\rangle(\cos\theta_3|H\rangle +
\sin\theta_3|V\rangle)+ |VV\rangle (\cos\theta_3|V\rangle+
\sin\theta_3|H\rangle)]$. We measure the correlation functions $E(Z_2,Z_3)$,
$E(X_1,X_2)$ under $|\psi_{123}\rangle$ where $\{|H\rangle,|V\rangle\}$,
$\{|H\rangle\pm|V\rangle\}$ are eigen bases for $Z$, $X$. Taking the
measured values in Eq. (\ref{corr-E}), the validity of the MDRs will be
verified [Fig. \ref{Fig-MDRs-qubit-result}(a)].

\section{Conclusions}

In this work, we have shown that the strength of correlation, which cab be
shared with a third particle through its interaction with one of the
particles in the entangled bipartite system, is not determined by the
interaction employed but by the fundamental measurement principle of QM. In
this sense, the multipartite nonlocality may be considered to be the
physical consequence of the uncertainty principle when quantum measurement
is involved, and hence, the essential elements of the quantum theory, i.e.,
the quantum entanglement, quantum nonlocality, and uncertainty principle,
are distinctly correlated in our scheme. This is heuristic and implies that
these essential elements should and could be investigated jointly. For
instance, the limit on measurement in quantum metrology
\cite{Quantum-metrology-N} may be modified by taking into account the
multipartite entanglement and MDR; the intricate correlation structures in
multipartite entanglement \cite{LU-2012}, on the other hand, indicate that
more investigations on the unexplored features of quantum measurement are
necessary. In order to ascertain the exact form of MDR through measuring the
correlation function, an optical experimental scheme has been proposed.
Finally, it should mentioned that although the analysis of measurement
precision and disturbance in this work is based on the definitions of
(\ref{def-precision}) and (\ref{def-disturbance}), it is also applicable to
other operator-type definitions.

\vspace{0.1cm}
\noindent {\Large \bf Acknowledgments}

\noindent This work was supported in part by the National Natural Science
Foundation of China (NSFC) under Grants No. 11121092, No. 11175249, No.
11375200, and No. 11205239.

\newpage

\appendix{\bf\Large Appendix}

\section{Functions $f_{\mathrm{q}}(\Delta A,\Delta B, \langle C\rangle)$}

For the Heisenberg-type MDR $\epsilon(A)\eta(B)\geq |\langle C\rangle|$, we
know that the shortest distance from the AR to the origin of the coordinate
is
\begin{eqnarray}
f_{\mathrm{He}}(\langle C\rangle) = 2 |\langle C\rangle| \; .
\nonumber
\end{eqnarray}
Ozawa's MDR [Eq. (5)] may be rewritten as
\begin{eqnarray}
[\epsilon(A) + \Delta A][\eta(B) + \Delta B] \geq \Delta A\Delta B
+ |\langle C\rangle| \; . \nonumber
\end{eqnarray}
The shortest distance from the AR to the origin for this MDR,
$f_{\mathrm{Oz}}$, may be solved by minimizing the value
$\sqrt{\epsilon(A)^2+\eta(B)^2}$ under the constraints of $[\epsilon(A) +
\Delta A][\eta(B) + \Delta B] \geq \Delta A\Delta B + |\langle C\rangle|$
and $\Delta A\Delta B\geq |\langle C\rangle|$. It is found that when $\Delta
A = \Delta B = \sqrt{|\langle C\rangle|}$ (i.e., the ideal minimum
uncertainty states for $A$, $B$), $f_{\mathrm{Oz}}$ gets its maximum value
of $(2-\sqrt{2})^2 |\langle C\rangle|$.

As expressed in Eq. (12), Branciard's MDR in Eq. (8) is an ellipse centered
at the origin. The minimal distance from the AR to the origin for this MDR
is equal to the minor axis of the ellipse, which is
\begin{eqnarray}
f_{\mathrm{B1}}(\Delta A,\Delta B,
\langle C\rangle) = \frac{1}{2}[\Delta A^2 + \Delta B^2 -
\sqrt{(\Delta A + \Delta B)^2 - 4\langle C\rangle^2 }] \; .\nonumber
\end{eqnarray}
When $\Delta A = \Delta B = \sqrt{|\langle C\rangle|}$, $f_{\mathrm{B1}}$
gets the maximum value of $|\langle C\rangle|$.

Along the same line, the corresponding expressions of $f_{\mathrm{q}}$ can
all be solved for other MDRs. We get, after numerical evaluations, that the
maximum attainable value from $r_{\mathrm{B2}}^2 = f_{\mathrm{B2}}$ form Eq.
(9) is $(4-2\sqrt{2}) |\langle C\rangle|$ (at $\Delta A =\Delta B
=\sqrt{|\langle C\rangle|}$). The values of Eqs. (6) and (7) may also be
obtained under their optimal measurements, i.e., $(\Delta A)^2 = (\Delta
\mathcal{A})^2 + \epsilon(A)^2$ \cite{Hall-2004}, where $r_{\text{Ha}}^2 =
2|\langle C\rangle|/5$ and $r^2_{\mathrm{We}} \approx 0.591|\langle
C\rangle|$ when $\Delta A = \Delta B = \sqrt{|\langle C\rangle|}$.

\section{The generality of Theorem \ref{Theorem-MDR-Correlation}}

Here we explicitly demonstrate that for all tripartite entangled states,
there is a constraint on correlations in the form of Eq. (\ref{Corr-ineq-3})
derived from the MDRs.

First, for arbitrary incompatible observables $A$ and $B$, Eq.
(\ref{Corr-ineq-3}) is satisfied by the set of quantum states $\Psi =
\{|\psi_{123}\rangle|\ |\psi_{123}\rangle = U_{13}|\psi_{12}\rangle
|\phi_3\rangle, U_{13}^{\dag}U_{13}=I\}$. All their local unitary equivalent
states $|\Psi_{123}\rangle = U_1\otimes U_2\otimes U_3|\psi_{123}\rangle \in
\Psi$ because
\begin{eqnarray}
|\Psi_{123}\rangle & = & U_{1}\otimes U_{2}\otimes U_{3}|\psi_{123}\rangle \nonumber \\
& = & U_{2} \otimes U_{13}'|\psi_{12}\rangle|\phi_3\rangle
= U_{13}'V_{1}|\psi_{12}\rangle|\phi_3\rangle \nonumber \\
& = & U_{13}''|\psi_{12}\rangle|\phi_3\rangle, \nonumber
\end{eqnarray}
where $U_{13}' = (U_1\otimes U_3) U_{13}$, $U_{13}''=U_{13}' (V_1\otimes
I_3)$, and we have used the fact that the unitaries $U_{2}|\psi_{12}\rangle
= V_{1}|\psi_{12}\rangle$ always exist for the quantum state
$|\psi_{12}\rangle = \frac{1}{\sqrt{N}} \sum_{i}|\alpha_i\rangle
|\alpha_i'\rangle$.

On the other hand, an arbitrary invertible operator $\Lambda$ makes the
transformation $|\widetilde{\psi}_{12}\rangle = \Lambda_2 |\psi_{12}\rangle$
by acting on particle 2. A set of states $|\widetilde{\psi}^{(i)}_1\rangle
=\ _2\langle q_i|\widetilde{\psi}_{12}\rangle/|_2\langle q_i
|\widetilde{\psi}_{12}\rangle|$ is prepared via a complete projection basis
$\{|q_i\rangle\}$. The measurement precision and disturbance for such states
are
\begin{eqnarray}
\tilde{\epsilon}^{\,(i)}(A)^2 = \langle \phi_3| \langle \widetilde{\psi}_1^{(i)}|
[ \mathcal{A} - A_1 \otimes I_3]^2
|\widetilde{\psi}_1^{(i)}\rangle |\phi_3\rangle \; ,  \nonumber \\
\tilde{\eta}^{\,(i)}(B)^2 = \langle \phi_3| \langle \widetilde{\psi}_1^{(i)}|
[\mathcal{B}- B_1\otimes I_3]^2|\widetilde{\psi}_1^{(i)}\rangle |\phi_3\rangle \; ,
\nonumber
\end{eqnarray}
where the tilde indicates the precision and disturbance are evaluated under
the states projected from $|\widetilde{\psi}_{12}\rangle$ and $\mathcal{A}$
and $\mathcal{B}$ have been defined in Eqs. (\ref{def-precision}) and
(\ref{def-disturbance}). Summing over the complete basis $\{|q_i\rangle\}$,
\begin{eqnarray}
\sum_{i} |_2\langle q_i
|\widetilde{\psi}_{12}\rangle|^2 \tilde{\epsilon}^{\,(i)}(A)^2 = \langle \phi_3|
\langle \widetilde{\psi}_{12}|
(U_{13}^{\dag} A_3 U_{13} - A_1)^2
|\widetilde{\psi}_{12}\rangle |\phi_3\rangle \; , \nonumber \\
\sum_i |_2\langle q_i
|\widetilde{\psi}_{12}\rangle|^2 \tilde{\eta}^{\,(i)}(B)^2 = \langle \phi_3|
\langle \widetilde{\psi}_{12}| (U_{13}^{\dag} B_1 U_{13}
- B_1)^2 |\widetilde{\psi}_{12}\rangle |\phi_3\rangle\; . \nonumber
\end{eqnarray}
Here, taking the measurement precision as an example, we have
\begin{eqnarray}
\sum_{i} |_2\langle q_i
|\widetilde{\psi}_{12}\rangle|^2 \tilde{\epsilon}^{\,(i)}(A)^2
& = & \langle \phi_3|\langle \widetilde{\psi}_{12}|
(U_{13}^{\dag}A_3^2 U_{13} + A_1^2 )
|\widetilde{\psi}_{12}\rangle |\phi_3\rangle - \nonumber \\
& &  \langle \phi_3|\langle \widetilde{\psi}_{12}|
(U_{13}^{\dag}A_3 U_{13} A_1 + A_1 U_{13}^{\dag} A_3 U_{13})
|\widetilde{\psi}_{12}\rangle |\phi_3\rangle \nonumber \\
& = & \langle \psi_{123}| A_3^2 \Lambda_2^{\dag}\Lambda_2 |\psi_{123}\rangle
+ \langle \psi_{123}| A_2'^2 \Lambda_2^{\dag}\Lambda_2 |\psi_{123}\rangle - \nonumber \\
& & \langle \psi_{123}| A_3( \Lambda_2^{\dag}\Lambda_2A_2' +
A_2'\Lambda_2^{\dag}\Lambda_2) | \psi_{123}\rangle \; , \nonumber
\end{eqnarray}
where $|\psi_{123}\rangle = U_{13}|\psi_{12}\rangle |\phi_3\rangle$. Since
$\langle \psi_{123}|A_2'^2\Lambda^{\dag}_2\Lambda_2 |\psi_{123}\rangle =
\langle \psi_{123}|\Lambda^{\dag}_2\Lambda_2 A_2'^2|\psi_{123}\rangle$,
\begin{eqnarray}
2 \sum_{i} |_2\langle q_i |\widetilde{\psi}_{12}\rangle|^2
\tilde{\epsilon}^{\, (i)}(A)^2
& = & \langle \psi_{123}|A_3^2 \Lambda_2^{\dag}\Lambda_2 +
\Lambda_2^{\dag}\Lambda_2 A_3^2 + A_2'^2\Lambda_2^{\dag}\Lambda_2 +
\Lambda_2^{\dag}\Lambda_2 A_2'^2 \nonumber \\
& & \hspace{1cm} -2A_3A_2' \Lambda_2^{\dag}\Lambda_2 - 2\Lambda_2^{\dag}\Lambda_2A_3A_2'
|\psi_{123}\rangle \; . \nonumber
\end{eqnarray}
Along the same line,
\begin{eqnarray}
2 \sum_i |_2\langle q_i |\widetilde{\psi}_{12}\rangle|^2 \tilde{\eta}^{\,(i)}(B)^2
& = & \langle \psi_{123}| B_1^2 \Lambda_2^{\dag}\Lambda_2 + \Lambda_2^{\dag}\Lambda_2 B_1^2
+B_2'^2\Lambda^{\dag}_2\Lambda_2 + \Lambda_2^{\dag}\Lambda_2B_2'^2 \nonumber \\
& & \hspace{1cm} - 2 B_1B_2'\Lambda_2^{\dag}\Lambda_2-
2\Lambda_2^{\dag}\Lambda_2B_1B_2' |\psi_{123}\rangle \; .\nonumber
\end{eqnarray}
Summing over the above two equations, defining $\widetilde{F}_{\mathrm{q}}
\equiv \sum_{i} |_2\langle q_i |\widetilde{\psi}_{12}\rangle|^2
\tilde{f}_{\mathrm{q}}^{(i)}$, where $\tilde{f}_{\mathrm{q}}^{(i)} \leq
\tilde{\epsilon}^{\,(i)}(A)^2 + \tilde{\eta}^{\,(i)}(B)^2$ are the same
function as that of Eq. (\ref{App-radius}) but evaluated under states
$|\widetilde{\psi}_{1}^{(i)}\rangle$, we have
\begin{eqnarray}
\widetilde{F}_{\mathrm{q}}^2 & \leq & |\langle \psi_{123} |[(A_3-A_2')^2+
(B_1-B_2')^2] \Lambda_2^{\dag}\Lambda_2 | \psi_{123} \rangle|^2 \nonumber \\
& \leq & \langle \psi_{123} | \Lambda_2^{\dag}\Lambda_2 [(A_3-A_2')^2+
(B_1-B_2')^2]\Lambda_2^{\dag}\Lambda_2 |\psi_{123} \rangle \nonumber \\
& & \times \langle \psi_{123} |[(A_3-A_2')^2 +
(B_1-B_2')^2] |\psi_{123} \rangle \; , \label{App-gen-derive}
\end{eqnarray}
where the equality holds when $\Lambda^{\dag}\Lambda=\lambda I$ in the
second inequality. The right-hand side of Eq. (\ref{App-gen-derive}) does
not depend on the choice of $\{|q_i\rangle\}$; therefore,
\begin{eqnarray}
\langle \widetilde{\psi}_{123}| (A_3-A_2')^2+
(B_1-B_2')^2| \widetilde{\psi}_{123} \rangle \xi \geq
\frac{\widetilde{\gamma}^{\,2}_{\mathrm{q}}}{\mathcal{N}}
\; , \label{App-gen-theorem}
\end{eqnarray}
where $|\widetilde{\psi}_{123}\rangle =
\Lambda^{\dag}_2\Lambda_2|\psi_{123}\rangle/\sqrt{\mathcal{N}}$, the
normalization factor $\mathcal{N} = \langle \psi_{123}|
(\Lambda_2^{\dag}\Lambda_2)^2 |\psi_{123}\rangle$, and
$\widetilde{\gamma}_{\mathrm{q}} = \mathrm{Max}\{ \widetilde{F}_{\mathrm{q}}
\}$ over $\{|q_i\rangle\}$, $\xi =\langle \psi_{123}| (A_3-A_2')^2+
(B_1-B_2')^2|\psi_{123} \rangle$. However, according to Eq.
(\ref{app-sum-radii}), $\xi= \sum_{i}|\ _2\langle p_i|\psi_{12}\rangle|^2
[\epsilon^{(i)}(A)^2 + \eta^{(i)}(B)^2]$ does not depend on the matrix
$\Lambda$. For varying $\Lambda$, $\xi$ may be chosen as a constant whose
value is $\gamma_{\mathrm{q}}$ which is determined by the fact that Eq.
(\ref{App-gen-theorem}) should reduce to Eq. (\ref{quadratic-inequality})
when $\Lambda^{\dag}\Lambda = I$. Therefore, Eq. (\ref{App-gen-theorem}) may
be rewritten as
\begin{eqnarray}
\langle A_2'A_3\rangle + \langle B_2'B_1\rangle \leq
\frac{1}{2}[\langle A_2'^2\rangle + \langle A_3^2\rangle +
\langle B_2'^2\rangle + \langle B_1'^2\rangle -
(\widetilde{\gamma}_{\mathrm{q}}^{\,2})/(\gamma_{\mathrm{q}}\mathcal{N}) ]
\; , \label{App-gen-corr}
\end{eqnarray}
which is a fundamental constraint on the bipartite correlations for the
quantum state $|\widetilde{\psi}_{123}\rangle$ and is the same as that of
Eq. (\ref{Corr-ineq-3}) for $|\psi_{123}\rangle$ due to the MDR.

The set of states satisfying Eq. (\ref{App-gen-corr}) may be formulated as
(up to a normalization)
\begin{eqnarray}
\widetilde{\Psi} = \{ |\widetilde{\psi}_{123}\rangle| \
|\widetilde{\psi}_{123}\rangle = \Lambda_2^{\dag}\Lambda_2 |\psi_{123}\rangle,
\mathrm{Det}[\Lambda]\neq 0, |\psi_{123}\rangle \in \Psi \}\; .
\end{eqnarray}
Defining the Hermitian operator as $ \Lambda^{\dag}\Lambda = H$,
$|\widetilde{\psi}_{123}\rangle$ may be generally expressed as
\begin{eqnarray}
|\widetilde{\psi}_{123}\rangle & = & \frac{1}{\sqrt{N}} \sum_{i} H| \alpha_i'\rangle_2
U_{13}U^{\dag}_3U_3\left(|\alpha_i\rangle_1\sum_{j=1}^N \gamma_j|\alpha_j\rangle_3\right) \nonumber \\
& = & \frac{1}{\sqrt{N}} \sum_{i}H | \alpha_i'\rangle_2
U_{13}U_{3}^{\dag}\left(|\alpha_i\rangle_1|\alpha_1\rangle_3\right) \nonumber \\
& = & \frac{1}{\sqrt{N}} \sum_{l,m,n,i} (|\alpha_l\rangle_1\langle \alpha_l|\otimes
|\alpha_m'\rangle_2\langle \alpha_m'| \otimes |\alpha_n\rangle_3\langle \alpha_n|)
H | \alpha_i'\rangle_2 U_{13}'\left(|\alpha_i\rangle_1|\alpha_1\rangle_3\right) \nonumber \\
& = & \frac{1}{\sqrt{N}} \sum_{l,m,n,i} h_{mi}\psi_{lin} |\alpha_l\rangle_1
|\alpha_{m}'\rangle_2 |\alpha_{n} \rangle_3 \; , \label{App-gen-123}
\end{eqnarray}
where we have used $|\phi_3\rangle = \sum_{j}\gamma_j|\alpha_j\rangle$, $U_3
|\phi_3\rangle =  |\alpha_1\rangle$; $U_{13}'=U_{13}U^{\dag}_3$ is another
unitary interaction matrix of $N^2\times N^2$, $h_{mi} =\ _2\langle
\alpha_m|H|\alpha_i\rangle_2$ , and $\psi_{lin}=u'_{ln,i1}=\
_1\langle\alpha_l| _3\langle \alpha_n | U'_{13}
|\alpha_i\rangle_1|\alpha_1\rangle_3$. The number of free real parameters in
$|\widetilde{\psi}_{123}\rangle$ includes $2N^3-N-N(N-1)$ real parameters
from $u'_{ln,i1}$ and $N^2$ from the Hermitian operator $H$. The total
number of $2N^3$ equals the number of real parameters of the quantum state
of $(N\times N\times N)$-dimensional tripartite states. Thus, for all the
tripartite states, Eq. (\ref{Corr-ineq-3}) or Eq. (\ref{App-gen-corr}) is
satisfied.

\section{Maximal value of $E(Z_2, Z_3) + E(X_1,
X_2)$ for three-qubit states}

An arbitrary three-qubit state may be expressed as
\begin{eqnarray}
|\varphi_{123}\rangle & = & a_1 |+++\rangle +  a_2 |++-\rangle +
a_3 |+-+\rangle + a_4 |+--\rangle \nonumber \\
& & + a_5 |-++\rangle +  a_6 |-+-\rangle +
a_7 |--+\rangle + a_8 |---\rangle \; .\nonumber
\end{eqnarray}
Here $a_i\in \mathbb{C}$, and the normalization requires
$\sum_{i=1}^{8}|a_i|^2 = 1$. The QM prediction of $E(Z_2, Z_3) + E(X_1,
X_2)$ for this arbitrary state takes the following form:
\begin{eqnarray}
E(Z_2, Z_3) + E(X_1, X_2)
& = & \langle \varphi_{123} |I_1\otimes Z_2\otimes Z_3 |\varphi_{123}\rangle+
\langle \varphi_{123}|X_1\otimes X_2\otimes I_3 | \varphi_{123} \rangle \nonumber \\
& = & |a_1|^2 + |a_4|^2 + |a_5|^2 + |a_8|^2
- |a_2|^2 - |a_3|^2 - |a_6|^2 - |a_7|^2 + \nonumber \\
& & a_1^*a_7 + a_2^*a_8 + a_3^*a_5 + a_4^*a_6 +
a_1a_7^* + a_2a_8^* + a_3a_5^* + a_4a_6^* \nonumber \\
& = & |\vec{r}_1|^2 - |\vec{r}_2|^2 + \vec{r}\,^*_1\cdot\vec{r}_2
+ \vec{r}_1\cdot\vec{r}\,^*_2 \; ,\nonumber
\end{eqnarray}
where $\vec{r}_1 \equiv \{a_1,a_4,a_5,a_8\}^{\mathrm{T}} $, $\vec{r}_2
\equiv \{a_7,a_6,a_3,a_2\}^{\mathrm{T}}$ and $|\vec{r}_1|^2 + |\vec{r}_2|^2
= 1$. We may set $|\vec{r}_1| = \cos\theta$, $|\vec{r}_2|=\sin\theta$, and
according to the Cauchy-Schwarz inequality $|\vec{r}\,^*_1\cdot\vec{r}_2|
\leq |\vec{r}_1||\vec{r}_2|$, we have
\begin{eqnarray}
|\vec{r}_1|^2 - |\vec{r}_2|^2 + \vec{r}\,^*_1\cdot\vec{r}_2 + \vec{r}_1\cdot\vec{r}\,^*_2
& = & |\vec{r}_1|^2 - |\vec{r}_2|^2 + 2 |\vec{r}\,^*_1\cdot\vec{r}_2|\cos\phi \nonumber \\
& \leq & \cos(2\theta) + \sin(2\theta)\cos\phi \leq \sqrt{2} \; .
\end{eqnarray}
Here $|\vec{r}\,^*_1\cdot\vec{r}_2|\cos\phi$ are the real part of the
complex number $\vec{r}\,^*_1\cdot\vec{r}_2$.

\end{document}